\documentclass[conference,compsoc]{IEEEtran}

\ifCLASSOPTIONcompsoc
  \usepackage[nocompress]{cite}
\else
  \usepackage{cite}
\fi

\usepackage[pdftex]{graphicx}
\graphicspath{{./figures/}}
\DeclareGraphicsExtensions{.pdf,.jpeg,.png}

\usepackage{amsmath}
\usepackage{algpseudocode}
\algnewcommand{\LineComment}[1]{\State \(\triangleright\) #1}

\usepackage{array}
\ifCLASSOPTIONcompsoc
  \usepackage[caption=false,font=footnotesize,labelfont=sf,textfont=sf]{subfig}
\else
  \usepackage[caption=false,font=footnotesize]{subfig}
\fi

\usepackage{url}
\usepackage{color,soul}

\begin{document}
\title{Using Hierarchical Parallelism to Accelerate the Solution of Many Small Partial Differential Equations}
\author{\IEEEauthorblockN{Jacob Merson}
\IEEEauthorblockA{Department of Mechanical,\\ Aerospace\\and Nuclear Engineering\\
Rensselaer Polytechnic Institute\\
110 Eighth St. Troy, NY 12180\\
Email: mersoj@rpi.edu}
\and
\IEEEauthorblockN{Mark S. Shephard}
\IEEEauthorblockA{Scientific Computation\\Research Center\\
Rensselaer Polytechnic Institute\\
110 Eighth St. Troy, NY 12180\\
Email: shephard@rpi.edu}
}

\maketitle

\begin{abstract}
%Kokkos hierarchical parallelism was leveraged to improve the performance of running many small explicit finite element problems simultaneously on a single NVIDIA Volta V100 GPU. This method was compared against a naive loop based approach, and a NVIDIA Multi-Process Service (MPS) parallel approach.
This paper presents efforts to improve the hierarchical parallelism of a two scale simulation code. Two methods to improve the GPU parallel performance were developed and compared. The first used the NVIDIA  Multi-Process  Service and the second moved the entire sub-problem loop into a single kernel using Kokkos hierarchical parallelism and a PackedView data structure. Both approaches improved parallel performance with the second method providing the greatest improvements. 

\end{abstract}

% Paper discussing running lots of ODE on GPU\cite{niemeyerGPUBasedParallelIntegration2014}

\section{Introduction}
Hierarchical multiscale methods are commonly used to model engineering materials which exhibit complex micromechanical behavior that is not easily captured with standard constitutive modeling \cite{feyelFE2MultiscaleApproach2000,smitPredictionMechanicalBehavior1998,mieheComputationalMicrotomacroTransitions2003,kanouteMultiscaleMethodsComposites2009}. This behavior is often caused by a material having a makeup of discrete constituents such as atoms, molecules, fibers, etc. For a two scale analysis the macroscale, or engineering scale, partial differential equations are often discretized by finite element methods. At each material point in the macroscale, a microscale sub-problem made from discrete components is solved to obtain the material constitutive properties at that point. This information passing scheme is often referred to as the FE2 scheme drawing from the fact that there are is a finite element analysis occurring on two scales \cite{feyelFE2MultiscaleApproach2000}.

In our multiscale implementation we use parallelism at multiple levels. On the macroscale, we use domain level parallelism to break up our finite element mesh. This is implemented using the SCOREC parallel unstructured meshing infrastructure (PUMI) \cite{ibanezPUMIParallelUnstructured2016}. Each microscale sub-problem is independent which leads to an ``embarrassingly parallel'' algorithm. To couple the scales, we use the Adaptive Multiscale Simulation Infrastructure which breaks processors on the target platform into an independent processor set for each scale \cite{tobinAdaptiveMultiscaleSimulation2018a}. In our previous work, the individual microscale sub-problems were not parallelized. Due to our increased understanding of the sub-scale physics in our problem of interest, we have increased the number of degrees of freedom in the microscale sub-problems by two to three orders of magnitude \cite{shahsavariSizeEffectMechanical2013,mersonSizeEffectsRandom2020}. This increase in problem size makes parallelization of the individual sub-problems essential to performing analyses with physical relevance.

Due to the increased computational cost associated with changes in the solution method described in section \ref{sec:multiscale_method} and the increase in microscale problem size, the solution to the microscale problems became a performance bottleneck. Therefore, we ported our code to use GPU parallelism for the microscale problems. For our problem of interest, the microscale problems often have less than 100,000 degrees of freedom. This is not large enough to saturate the GPU with our current analysis methods which primarily consists of vector operations---similar to BLAS level 1 (See figure \ref{alg:dynamic_relaxation}).

To achieve adequate GPU throughput, multiple microscale problems must be solved on each GPU at a time. This was accomplished using two methods. The first was using NVIDIA Multi-Process Service (MPS) which unintrusively allows multiple processes to launch GPU kernels at a time \cite{nvidiaMultiProcessService2019}. Although MPS gave a good speedup, our analysis was still limited by kernel launch overhead and by the limited number of processors on each node to launch GPU kernels from. Additionally, great care must be taken to use MPS in an environment with multiple GPUs per node because improper MPS setup causes a drastic reduction in the weak scalability. The second method for increasing GPU throughput was to pack multiple microscale problems into a single kernel launch using the Kokkos hierarchical parallelism construct \cite{carteredwardsKokkosEnablingManycore2014}. A more thorough discussion of this method is given in section \ref{sec:parallel_implementation}. To aid in understanding the selection of our parallelization strategy, a more comprehensive discussion of multiscale modeling techniques is presented below.

\section{Multiscale Modeling of Biological Tissues} \label{sec:multiscale_method}
One particular application of these methods is to model biological tissues which are made of constituent collagen fiber networks. Typically, modeling biological tissues requires large strain analysis because they are soft and physiological strains can often exceed 50\%. The FE2 method has been extended to allow for large strains; however, the methods discussed in the literature utilize implicit finite element methods for both analysis scales \cite{mieheNumericalComputationAlgorithmic1996, smitPredictionMechanicalBehavior1998}. Unfortunately, the deformation of fiber networks is highly nonlinear and the network can go through bifurcation points, or may not be isostatic, i.e. the tangent stiffness matrix can be singular during an analysis \cite{licupElasticRegimesSubisostatic2016}. As a result, athermal fiber networks, such as collagen networks, are typically modeled with explicit finite element methods \cite{islamStochasticContinuumModel2018,deogekarStrengthRandomFiber2018}. The use of a purely explicit analysis for the sub-scale problems in a multiscale analysis is problematic because kinetic energy is lost in the microscale-to-macroscale coupling. Additionally, inertial effects can change the microscale material properties. To work around these issues, a dynamic relaxation method is used for the microscale problems. Dynamic relaxation works by mapping a static analysis to a damped dynamic explicit one where the system residual is monitored for convergence \cite{underwoodDynamicRelaxation1986}.

The use of the dynamic relaxation method at the microscale greatly improves the global strains which the multiscale method can achieve for fibrous materials, however it imposes a significant computational cost. Our previous studies have shown the multiscale analysis of biological tissues at scale using homogeneous computing technologies with MPI based parallelism \cite{tobinAdaptiveMultiscaleSimulation2018a,chanImagebasedMultiscaleMechanical2018}. Due to the increased computational cost of dynamic relaxation, physiologically relevant problems are no longer accessible through MPI based parallelism alone. Therefore, the microscale portions of our analysis code have been ported to use GPU accelerators. 

The variability in GPU programming environments across hardware vendors poses a significant challenge to the maintainability of a GPU accelerated code which must run on a variety of systems. Therefore, we chose to use the Kokkos C++ library for GPU support. Kokkos is a C++ programming model which is designed to enable performance portability \cite{carteredwardsKokkosEnablingManycore2014,KokkosPerformancePortabilitya}. The Kokkos team has committed to maintaining support for all of the vendors who are providing accelerators for the Department of Energy leadership class computing resources (AMD, Intel, NVIDIA). This support allows for writing a single version of the analysis code that will run across most of the easily accessible GPU accelerators.

\section{Parallel Implementation} \label{sec:parallel_implementation}

\begin{figure}
\caption{Dynamic relaxation algorithm}
\label{alg:dynamic_relaxation}
\begin{algorithmic}[1]
\State Load mesh and compute edge connectivity
\State Compute the mesh connectivity array
\State Transfer the connectivity array to the device (GPU)
\State Set displacement boundary conditions on fixed nodes (fixed dof vector operation)
\State Compute mass matrix (finite element integration)
%\State $\mathbf{v}\gets\mathbf{v^0}$, $\mathbf{u}\gets 0$, $n\gets0$, $t\gets0$, $r_c\gets 1E-6$\Comment{Initialize system}
\State \Call{getInternalForces}{$\mathbf{u}$}
\State update accelerations (free dof vector operation)
\Repeat
\State Compute next time step
\State Partial velocity update (free dof vector operation)
\State Update displacements (free dof vector operation)
\State \Call{getInternalForces}{$\mathbf{u}$}
\State Compute damping force (free dof vector operation)
\State Compute force residual (free dof vector reduction)
\State Update Accelerations (free dof vector operation)
\State Partial velocity update (free dof vector operation)
\State Optionally Check Energy Balance (3 vector reductions)
\State Update iteration count
\Until{Force residual converged}
\end{algorithmic}
\end{figure}

Figure \ref{alg:dynamic_relaxation} gives the basic dynamic relaxation algorithm we used for the microscale sub-problems. This algorithm is identical to a two step central difference method found in any finite element text book, with the exception that the convergence criteria is based on a force residual measurement rather than time. Note that each sub-scale problem converges at a different rate which can lead to load imbalance.

In the naive approach, this algorithm was carried out using fused kernels for any subsequent operations with the same loop characteristics. The benefits of kernel fusion have been discussed extensively in the literature both for the case of explicit ODEs, and general GPU computations \cite{korchAcceleratingExplicitODE2018,wangKernelFusionEffective2010,wahibScalableKernelFusion2014}. The \texttt{GetInternalForces} subroutine accounts for two Kernel launches: the first to zero the internal force vector, and the second to scatter the elemental internal forces to the nodes. The current implementation uses atomic operations to scatter the forces.

In this naive approach, a number of microscale sub-problems were assigned to each MPI rank, and were executed serially with respect to each other within each rank. Despite the use of GPU acceleration for the vector operations, this approach had poor performance for the sub-scale problems with small numbers of degrees of freedom when compared with a CPU-only implementation with serial vector operations. To unintrusively improve this naive approach, NVIDIA MPS was used to allow kernels from multiple MPI ranks to run concurrently. The use of MPS led to significant performance improvements for small DOF problems compared with the naive case. Problem size and number of simultaneous MPI ranks used with MPS can have a drastic effect on performance. All MPS results presented in section \ref{sec:results} use 32 MPI ranks per GPU which gives the best performance in the range of problem sizes discussed here.

Since the loop in algorithm \ref{alg:dynamic_relaxation} executes millions of times per macroscale simulation step, we observed that this approach had significant kernel launch overhead. To overcome this, we moved the entire loop into a single kernel. This was done using Kokkos hierarchical parallelism which uses teams of threads to enable a 2D map to the hardware. The CUDA reciprocal to this mechanism is launching a 1D grid of 1D blocks. Since our sub-scale problems each have less than 10,000 free degrees of freedom, we found that good performance could be achieved by assigning one thread team to each sub-scale problem. Here, we juxtapose the free degrees of freedom which are those without any Dirichlet constraints, to what we call degrees of freedom which are all potential degrees of freedom. Unlike an implicit FEM method, the constrained degrees of freedom cannot be completely eliminated as they are needed for the internal force computation. Reordering the fixed degrees of freedom to a contiguous block at the end of the displacement array allows most of the update algorithm to only operate on the smaller proportion of free degrees of freedom (figure \ref{alg:dynamic_relaxation}).

The choice of number of threads per team had a strong effect on performance. The ideal number of threads per team is a function of the microscale problem size. All presented results use 512 threads per team, which provided a good compromise for the performance of the smallest and largest systems we tested. 

A \texttt{PackedView} data structure which has similar semantics to a Kokkos \texttt{DualView} was used to allow effective access to N-D vector data within each thread team \cite{mersonKokkospackeddata}. This data structure uses a row vector and value vector, similar to those from compressed row storage (CRS), to store the data associated with all sub-scale problems on the current MPI rank in a contiguous array in memory. Each sub-scale problem gains access to the correct portion of memory through a Kokkos \texttt{Subview}. In some ways, this structure is similar to a Kokkos \texttt{View} of \texttt{Views}. However, with the current implementation the \texttt{PackedView} can not be resized after initialization. A comprehensive performance comparison between the \texttt{PackedView} data structure, and \texttt{View} of \texttt{Views} has not been performed to date. This differs from the \texttt{StaticCrsGraph} in Kokkos which cannot handle non-integral datatypes, and does not have \texttt{DualView} semantics.

Although moving the analysis loop inside of a single kernel launch was effective for our problems of interest, it can easily succumb to low performance from high register pressure. Significant effort had to be made to reduce the register pressure and ensure that multiple warps could be concurrently scheduled. One mechanism we used to reduce register pressure was to move some of the variables which are carried across loop iterations such as the pseudo-time and the loop iteration count into shared memory. We found that performance gain from the reduction in register pressure outweighed the loss in bandwidth from moving these variables to shared memory. The need to reduce register usage in this single kernel implementation led us to favor a stripped-down version of our algorithm which was specific to the physical system at hand. In other words, flexibility of our code had to be sacrificed to obtain improved performance characteristics.

\section{Results} \label{sec:results}

\begin{figure}
    \centering
    \includegraphics[width=\linewidth]{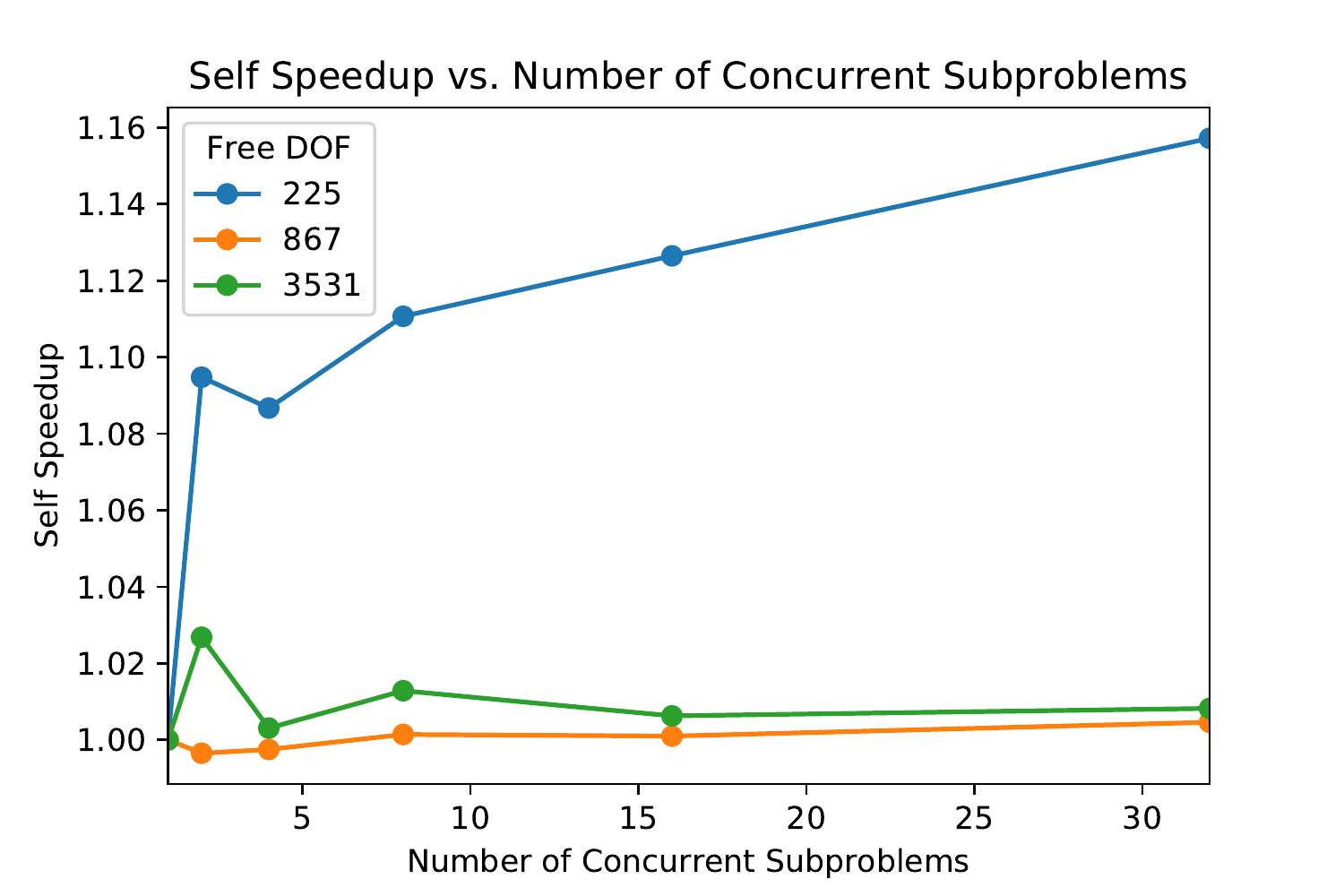}
    \caption{Speedup of the naive loop based analysis normalized by the number of concurrent sub-problems compared with the loop based single sub-problem. A flat line corresponds to a linear increase in runtime. Each data point is the mean of three analysis runs.}
    \label{fig:loop_self_speedup}
\end{figure}

\begin{figure}
    \centering
    \includegraphics[width=\linewidth]{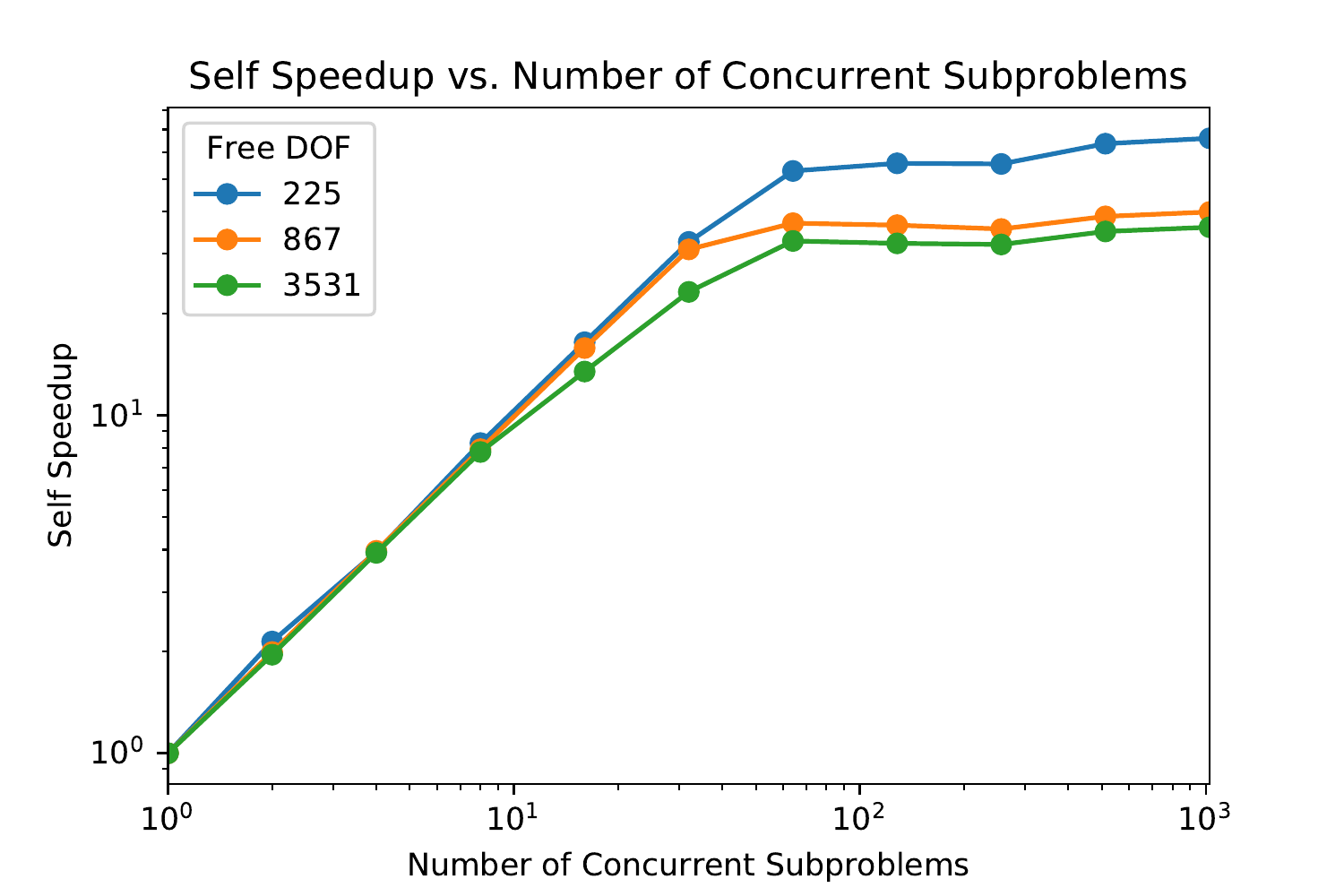}
    \caption{Speedup of the thread team based analysis normalized by the number of concurrent sub-problems compared with the thread team based single sub-problem. Each data point is the mean of three analysis runs.}
    \label{fig:team_self_speedup}
\end{figure}

The performance results presented here, are all computed on a single Volta V100 GPU---part of an IBM AC922 node. Each AC922 node contains 2, 20 core IBM power 9 processors clocked at 3.15GHz, 512 GiB of RAM, and 6 Volta V100 GPUs. The code is compiled with version 16.1.0 of IBM's XL compiler for host code, version 10.1 of Cuda, and version 3.1 of Kokkos. The MPS results make use of Spectrum MPI version 10.3.

Figures \ref{fig:loop_self_speedup} and \ref{fig:team_self_speedup} show the runtime of a single sub-problem divided by runtime normalized by the number of concurrent sub-problems. This gives a measure of the speedup of a single sub-problem when computed in a concurrent batch. Since we are using the analysis technique's own single sub-problem runtime as a baseline for the speedup, we call this the ``self speedup''. For the naive loop based case (figure \ref{fig:loop_self_speedup}), we see the self speedup is very flat which indicates the expected linear increase in runtime. The smallest problem size sees a slight self speedup. When thread team based parallelization is used, a significant self speedup is observed (figure \ref{fig:team_self_speedup}). Here we see a initial regime of linear self speedup and a plateau regime for large numbers of concurrent sub-problems. In this initial linear scaling regime, the runtime remains flat since the numerical workload is not large enough to overcome the kernel launch latency. Interestingly, the initial self speedup is almost identical for each of the problem sizes we tried. The plateau region show that as the number of degrees of freedom in the problem increase, the self speedup decreases.

\begin{figure}
    \centering
    \includegraphics[width=\linewidth]{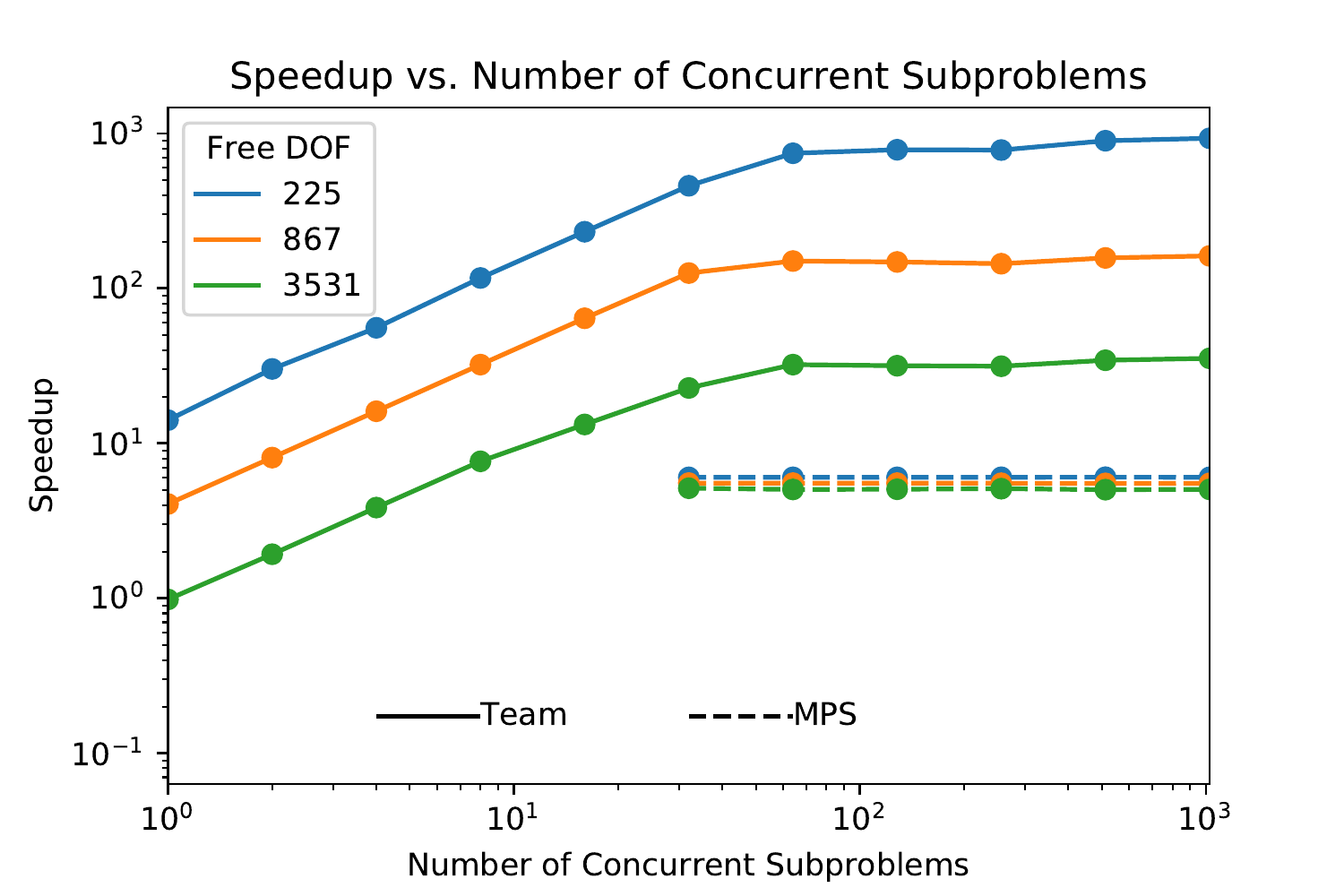}
    \caption{Speedup of the team based (solid line) and MPS based (dashed line) analysis over the naive approach. The MPS lines correspond to launching GPU kernels from 32 MPI ranks simultaneously. Each data point is the mean of three analysis runs.}
    \label{fig:speedup}
\end{figure}

Figure \ref{fig:speedup} shows the speedup of the thread team based and MPS based analysis methods over the naive loop based approach. In this plot, we see an initial linear scaling and a plateau region. We observe that as the problem size increases, the speedup obtained from the team thread based method decreases. This is likely due to a reduction in percentage of the problem which resides in the cache. We also observe that MPS based parallelism does provide some speedup over the naive approach, but it is not as effective as the thread team based approach. Also, for MPS, the speedup plateau does not depend on the problem size. One way to interpret these results is that for maximum efficiency, at least 80 concurrent sub-problems should be run on each GPU. Since the V100 has 80 SMs (streaming multiprocessors), this is consistent with each Kokkos team (CUDA block) occupying a single SM.

The speedups achieved using thread team parallelism tell a compelling story that moving an analysis loop inside of a single heavy weight kernel can be an effective optimization mechanism for problems that need to solve many problems which cannot saturate the GPU on their own. Although MPS seemed like it might be a reasonable solution, it suffered from still incurring a high kernel call latency due to the many kernels which were being called inside of a hot loop. Additionally, the MPS solution was not able to make as effective use of the cache since many different sub-scale problems were competing to be scheduled simultaneously, and each sub-scale problem that was scheduled in an interleaved fashion would cause cache misses.

\ifCLASSOPTIONcompsoc
  \section*{Acknowledgments}
\else
  \section*{Acknowledgment}
\fi
This work was supported in part by the National Institutes of Health (NIH) through Grant No. U01 AT010326-06. Also, this material is based upon work supported by the National Science Foundation Graduate Research Fellowship under Grant No. DGE-1744655.



\bibliographystyle{IEEEtran}
\bibliography{IEEEabrv,HiPar-2020}

\end{document}